\documentclass[aps,prb,reprint,amsmath,amssymb,superscriptaddress,showpacs,floatfix]{revtex4-1}
\usepackage{graphicx}
\usepackage{dcolumn}
\usepackage{bm}
\usepackage[colorlinks, allcolors=blue]{hyperref}
\usepackage[usenames, dvipsnames]{color}
\usepackage{setspace}
\usepackage{scalerel}
\usepackage{stackengine}
\usepackage{comment}
\stackMath
\newsavebox\tmpbox

\begin{document}

\title{Quantum phases of a frustrated spin-1 system: The 5/7 skewed ladder}
\author{Sambunath Das}
\email{sambunath.das46@gmail.com}
\affiliation{Solid State and Structural Chemistry Unit, Indian Institute of Science, Bangalore 560012, India}

\author{Dayasindhu Dey}
\email{dayasindhu.dey@gmail.com}
\affiliation{Solid State and Structural Chemistry Unit, Indian Institute of Science, Bangalore 560012, India}

\author{Manoranjan Kumar}
\email{manoranjan.kumar@bose.res.in}
\affiliation{S. N. Bose National Centre for Basic Sciences, Block - JD, Sector - III, Salt Lake, Kolkata - 700106, India}

\author{S. Ramasesha}
\email{ramasesh@iisc.ac.in}
\affiliation{Solid State and Structural Chemistry Unit, Indian Institute of Science, Bangalore 560012, India}

\date{\today}

\begin{abstract}
	The quantum phases in a spin-1 skewed ladder system formed by alternately fusing
	five- and seven-membered rings are studied numerically using the exact diagonalization
	technique up to 16 spins and using the density matrix renormalization group method 
	for larger system sizes. The ladder has a fixed isotropic antiferromagnetic (AF) exchange
	interaction ($J_2 = 1$) between the nearest-neighbor spins along the legs and a varying
	isotropic AF exchange interaction ($J_1$) along the rungs. As a function of $J_1$,
	the system shows many interesting ground states (gs) which vary from different types
	of nonmagnetic and ferrimagnetic gs. The study of diverse gs 
	properties such as spin gap, spin-spin correlations, spin density and bond 
	order reveal that the system has four distinct phases, namely, the AF phase at small 
	$J_1$; the ferrimagnetic phase with gs spin $S_G = n$  for 
	$1.44 < J_1 < 4.74$ and with $S_G = 2n$ for $J_1 > 5.63$, where $n$ is the 
	number of unit cells; and a reentrant nonmagnetic phase at 
	$4.74 < J_1 < 5.44$. The system also shows the presence of spin current at specific 
	$J_1$ values due to simultaneous breaking of both reflection and spin 
	parity symmetries.
\end{abstract}

\maketitle

\section{\label{sec:intro}Introduction}
In low-dimensional magnetic systems, confinement leads to strong  quantum 
fluctuations, and these systems can show many exotic phases in the presence of
frustration induced by the topology of exchange interactions~\cite{ckm69a,*ckm69b,
hamada88,chubukov91,chitra95,white96,itoi2001,mahdavifar2008,sirker2010,mk2015,
soos-jpcm-2016,mk_bow,chubukov91,mk2012,mk2015,vekua2007,hikihara2008,sudan2009,
dmitriev2008,meisner2006,meisner2007,meisner2009,aslam_magnon,kecke2007}. 
Even in a one-dimensional (1D) spin system, with only a
nearest-neighbor Heisenberg antiferromagnetic (HAF) exchange interaction, the  
ground state (gs) can be  gapped or gapless for integer or half odd-integer spins,
respectively, as pointed out in a seminal paper by Haldane~\cite{haldane83a,*haldane83b}. 
The gs of the HAF integer spin chain can be represented as a valance bond solid 
(VBS)~\cite{aklt87,aklt88,schollwock96}, and  in 1987 Affleck, Kennedy, Lieb, and Tasaki (AKLT) showed that 
perfect VBS state may exist on various geometries  with specific spins~\cite{aklt87,aklt88}. The 
AKLT state still continues to inspire physicists for various reasons; for example, 
the AKLT state has led to many recent developments such as the matrix product 
states technique~\cite{rommer95,verstraete2008,schollwock2011} which is a form of the density 
matrix renormalization group (DMRG) method~\cite{white-prl92,white-prb93,schollwock2005,karen2006},  the tensor 
network method~\cite{orus2014} and the projected entangled pair states 
ansatz~\cite{verstraete2008,schollwock2011}.  
AKLT states can also be represented as cluster states which can be used in 
measurement-based quantum computation~\cite{cirac2004,affleck2011}, and recently these states have been
explored in a spin-3/2 on a hexagonal lattice~\cite{lemm20, pomata20}.

The HAF spin-1 chain exhibits a topological phase, spin-1/2 edge modes, and the 
gs is four fold degenerate in the thermodynamic limit. The 
correlation length in the gs of spin-1 is 6.05 lattice units and the eigenvalue 
spectrum has large spin gaps~\cite{dayaprb16,white-huse-prb93}. The gs can be 
represented as a VBS, which belongs to the same universality class of AKLT 
states~\cite{aklt87,aklt88}.  The two leg HAF spin-1 
ladder shows interesting properties like plaquette-singlet solid state, 
where two spin-1/2 singlet dimers are sitting at each rung and there is no overlap 
between the VBS states in the large rung exchange limit~\cite{todo2001}.  
The AKLT state in the system breaks down for any finite value of rung exchange 
interaction~\cite{todo2001}. The spin-1 zigzag ladder shows a transition from a
Haldane phase to a double Haldane phase~\cite{hikihara2002,natalia2016}. In fact the zigzag 
ladder can be mapped into a chain system with nearest-neighbor and next nearest-neighbor 
exchange interactions, and the gs of the frustrated systems is a singlet. 
In this work, we explore the magnetic phases of a spin-1 system on a 
5/7-skewed ladder system; it has been demonstrated that a spin-1/2 system on
this lattice shows many exotic phases~\cite{geet}.

The 5/7-skewed ladder is inspired by fused Azulene, a ladder like 
structure made up of 5- and 7-membered carbon rings alternately fused on a chain,
studied by Thomas {\emph et al.} in which they showed that the gs is 
ferrimagnetic~\cite{thomas2012}. These 
structures can be mapped to a zigzag like ladder structure with some missing 
bonds~\cite{thomas2012,geet}. The HAF spin-1/2 system on various lattices such as  the 5/7, 3/4, 3/5, and
5/5 is studied and it was shown that the gs of these systems exhibits
many interesting magnetic and nonmagnetic gs in their quantum phase 
diagrams with strength of the rung exchange interaction as a phase parameter~\cite{geet}. 
In the large rung exchange limit, the gs wavefunction of a 5/7 skewed ladder can be 
represented as a product of rung singlet dimers and two ferromagnetically interacting 
spins per unit cell~\cite{57plateau}.  In various parameter regimes this system shows dimer, 
spiral and chiral vector phases~\cite{geet}.  In the presence of an axial magnetic field the 
HAF spin-1/2 system on the 5/7 skewed ladder exhibits four magnetization plateau phases~\cite{57plateau}.

\begin{figure}
\includegraphics[width=3.4 in]{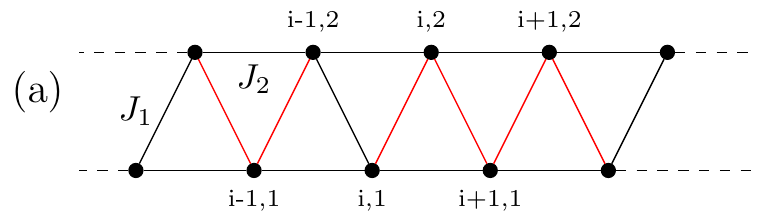}
\includegraphics[width=3.4 in]{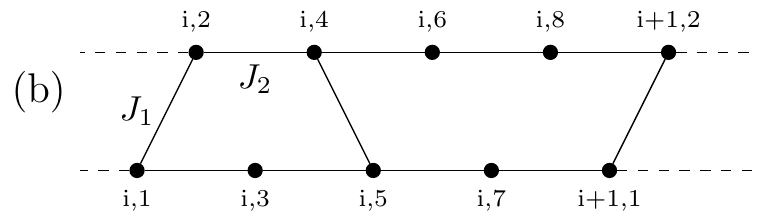}
\caption{\label{fig:schematic}Schematic diagram of (a) a zigzag ladder. The
nearest-neighbor or rung interaction is $J_1$ and the next-nearest neighbor 
(along the leg) interaction is $J_2$. (b) The 5/7 skewed ladder: Some rung bonds shown
in red of the zigzag ladder are periodically removed to give a 5/7 skewed ladder.
Here ``$i$'' is the index of the unit cell and the numerals 1, 2, \ldots are numbering
of the spins within the unit cell. There are 2 spins per unit cell in the zigzag
ladder while there are 8 spins per unit cell in the 5/7 ladder. The sites on the 
top leg are even numbered and on the bottom leg are odd numbered.}
\end{figure}

The structure of zigzag and 5/7 skewed ladder are shown in Figs.~\ref{fig:schematic}(a) 
and~\ref{fig:schematic}(b) and by periodically removing some of the rung bonds, shown in red, from 
Fig.~\ref{fig:schematic}(a) to give 5/7 skewed ladder in Fig.~\ref{fig:schematic}(b). 
In this paper, we are interested in the gs phases of a spin-1  5/7 skewed ladder  
as a function of the ratio of rung-to-leg exchanges $J_1$ and $J_2$, respectively. 
We show that this system is highly frustrated, and in the small rung interaction 
limit, $J_1 / J_2 < 1.06$,  singlet dimers along the rung are weak and 
correlations along the leg remain short ranged,  
whereas, for $J_1 / J_2 > 1.44$, the gs is magnetic and each unit cell contributes spin-1 
to the gs spin $S_G$, and spin densities are distributed over the whole unit cell, with 
spin density at sites 3 and 7 being large. For $4.74 < J_1 / J_2 < 5.44$ the system is 
nonmagnetic but for $J_1>5.63$ gs of the system is magnetic with each unit cell 
contributing spin 2 to $S_G$ with prominent rung dimers and site spin densities.

This paper is divided into four sections. In Section~\ref{sec2} we discuss the model 
Hamiltonian and the numerical methods. The results are presented and discussed in 
Section~\ref{sec3} under four subsections. Section~\ref{sec4} provides a summary  of results 
and conclusions.

\section{\label{sec2}Model and Method}
The site numbering used in this paper for the 5/7 skewed ladder is shown in
Fig.~\ref{fig:schematic}(b). All nonzero exchange interactions between spins are 
antiferromagnetic (AF). The sites are numbered such that odd numbered sites 
are on the bottom leg and even numbered sites are on the top leg. Thus the rung 
bonds are the nearest-neighbor exchanges $J_1$ and the bonds on the legs are 
the next-nearest-neighbor exchanges $J_2$. The exchange $J_2$ is set 
to 1  and it defines the energy scale. The model Hamiltonian of the 5/7 skewed ladder 
can be written as 
\begin{eqnarray}
	H_{5/7} &=& J_1 \sum_i \left(\vec{S}_{i,1} \cdot \vec{S}_{i,2} + \vec{S}_{i,4}
	\cdot \vec{S}_{i,5} \right)  \nonumber \\
	& &+ J_2 \sum_i \bigg( \vec{S}_{i,7} \cdot 
	\vec{S}_{i+1,1}
	+ \vec{S}_{i,8} \cdot \vec{S}_{i+1,2}  \nonumber \\
	& &  \qquad \qquad + \sum_{k=1}^6
	\vec{S}_{i,k} \cdot \vec{S}_{i,k+2} \bigg).
\label{eq:ham57}
\end{eqnarray}
where $i$ labels the unit cell and $k$ are the spins within the unit cell
(Fig.~\ref{fig:schematic}). The first term denotes the rung exchange terms, and the 
second term denotes the exchange interactions along the legs.

We use the exact diagonalization technique for finite ladders with up to 16
spins and impose periodic boundary condition (PBC). There are mirror planes 
perpendicular to the ladder, for example, the plane perpendicular to the ladder 
and passing through site 3 and the perpendicular bisector of sites 2 and 4, as 
well as the one passing through site 7 and the perpendicular bisector of sites
6 and 8, again perpendicular to the ladder. An extra rung is needed when the open 
boundary condition (OBC) is used. For larger 
system sizes we use the DMRG method~\cite{white-prl92,white-prb93,schollwock2005,karen2006}
to handle the large degrees of freedom in the many body Hamiltonian. We retain 
up to 500 block states ($m=500$), which are the eigenvectors of the block density
matrix with dominant eigenvalues. The chosen value of ``$m$'' keeps the 
truncation error to less than $\sim 10^{-10}$. We also carry out 6--10 finite 
sweeps for improved convergence. The details of building the 5/7 ladder for the 
DMRG method is the same as in Ref.[~\onlinecite{geet}]. The largest system size 
studied is a system with 130 sites or 16 unit cells with OBC. The
DMRG calculations are carried out for different $S^z$ values of ladders. The
gs spin is $S_G=l$, for $l$ that satisfies $\Gamma_l = 0$ and 
$\Gamma_{l+1} > 0$, where $\Gamma_l$ is given by
\begin{equation}
\Gamma_l = E_0 (S^z = l) - E_0 (S^z = 0),
\label{eq:gap}
\end{equation}
with $E_0$ being the lowest energy state in the chosen $S^z$ sector. The
correlation function and bond orders are computed in the gs, with 
$S^z = S$.

\section{\label{sec3}Results and discussions}
In the gs, the spin-1 5/7 skewed ladder, like the spin-1/2 system, also shows
many exotic phases like the bond order wave (BOW) phase, chiral points in
parameter space of the Hamiltonian  and
nonmagnetic to magnetic phase transition on tuning the value of
$J_1$. However, there are significant differences from the spin-1/2 system. To
analyze the magnetic transitions in the quantum phase diagram, various quantities
are analyzed as a function of $J_1 / J_2$ which is the only variable model parameter in this 
system and $J_2$ is set to 1. Besides the spin gaps $\Gamma_l$, we have the computed
correlation function $C(r) = \langle \vec{S}_i \cdot \vec{S}_{i+r} \rangle$ to study
the behavior of the spins in the system. The bond order between 
bonded neighbors $-\langle \vec{S}_{i} \cdot \vec{S}_{i'} \rangle$ where sites 
$i$ and $i'$ are bonded neighbors, and spin-density 
$\langle S^z_{i} \rangle$ within a unit cell are also calculated and compared 
with the results for a 1D spin-1 chain where appropriate.

\subsection{Nature of gs}
\begin{figure}
\includegraphics[width=3.4 in]{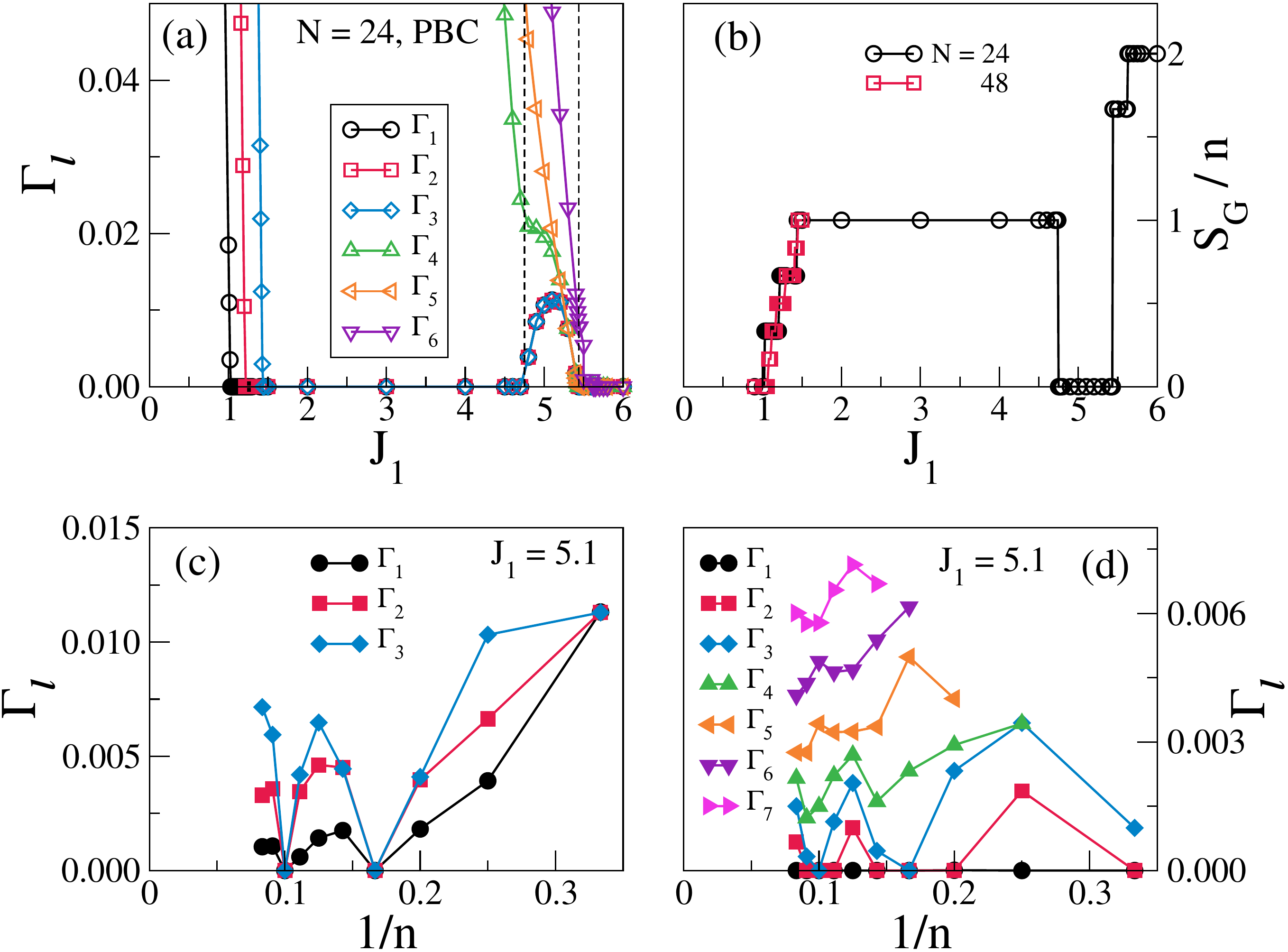}
	\caption{\label{fig:spgap}(a) The lowest excitation gaps $\Gamma_l$ for
different $S^z = l$ manifolds are shown as functions of $J_1$. For
$J_1 < 1.06$, $\Gamma_1$ is nonzero, whereas for $1.06 < J_1 < 4.74$,
$\Gamma_l$ is zero for $l \leq S$ where $S$ is the total spin of the gs. The
system exhibits reentrant nonmagnetic phase for $4.74 < J_1 < 5.44$ 
where $\Gamma_1$ is nonzero. For $1.44 < J_1 < 4.74$ $S_G \sim n$ and
	for $J_1 > 5.63$ $S_G \sim 2n$. (b) The variation of the gs
	spin per unit cell $S_G / n$ with $J_1$ for system sizes of
	48 $(n=6)$ spins in regions I and II, and of $24$ $(n=3)$ spins in all
	the regions. (c) The lowest excitation gaps $\Gamma_1$, $\Gamma_2$ and 
	$\Gamma_3$ with the inverse of the number of unit cells of a 5/7 with 
	PBC are shown for $J_1 = 5.1$. (d) The lowest excitation gaps $\Gamma_1$
	to $\Gamma_7$ with the inverse of the number of unit cells of a 5/7 with 
	OBC are shown for $J_1 = 5.1$.} 
\end{figure}

The spin in the gs, $S_G$, of the skewed 5/7 ladder systems is obtained 
from the magnetic gaps $\Gamma_l$ defined in Eq.~\ref{eq:gap}. In
Fig.~\ref{fig:spgap}(a), we plot the gaps $\Gamma_l$ for different values of ``$l$'', 
as a function of $J_1$ for a system with 24 spins corresponding to 3 unit cells $(n=3)$
under PBC. The plot shows that 
there are four distinct regions: In region I with $0 < J_1 < 1.06$, the gs
is a singlet and is nonmagnetic; in region II, $1.06 < J_1 < 4.74$, $S_G$ is 
less than or equal to the number of unit cells in the systems, consequently each unit cell 
contributes at most spin-1 to $S_G$ and for $1.44 < J_1 < 4.74$, the 
spin $S_G$ saturates to the number of unit cells. We calculate $S_G / n$ 
as a function of $J_1$ for systems with $N = 24$ and $N = 48$ spins to investigate 
if the transition to $S_G  = n$ is smooth or abrupt. We find that $S_G / n$ shows
a gradual increase in the region $1.06 < J_1 < 1.44$ for the larger system, shown 
in Fig.~\ref{fig:spgap}(b). The weak finite size effect in $S_G / n$ can be 
attributed to the short spin-spin correlation lengths. As shown in Fig.~\ref{fig:spgap}(b),
the increase in $S_G / n$ is continuous between regions I and II. In region III, 
$4.74 < J_1 < 5.44$, the gs becomes nonmagnetic for the 24 spin system. The 
spin-spin correlation function reveals that in this region the ``free'' spins in 
each unit cell align ferromagnetically while the alignment of the spins across 
unit cells is AF, with large periodicity. The transition from 
region II to region III is abrupt for $N = 24$ spins and we had convergence 
difficulties even for the $N = 48$ spins and hence cannot comment on the effect 
of system size. 

\begin{figure}
	\includegraphics[width=3.4 in]{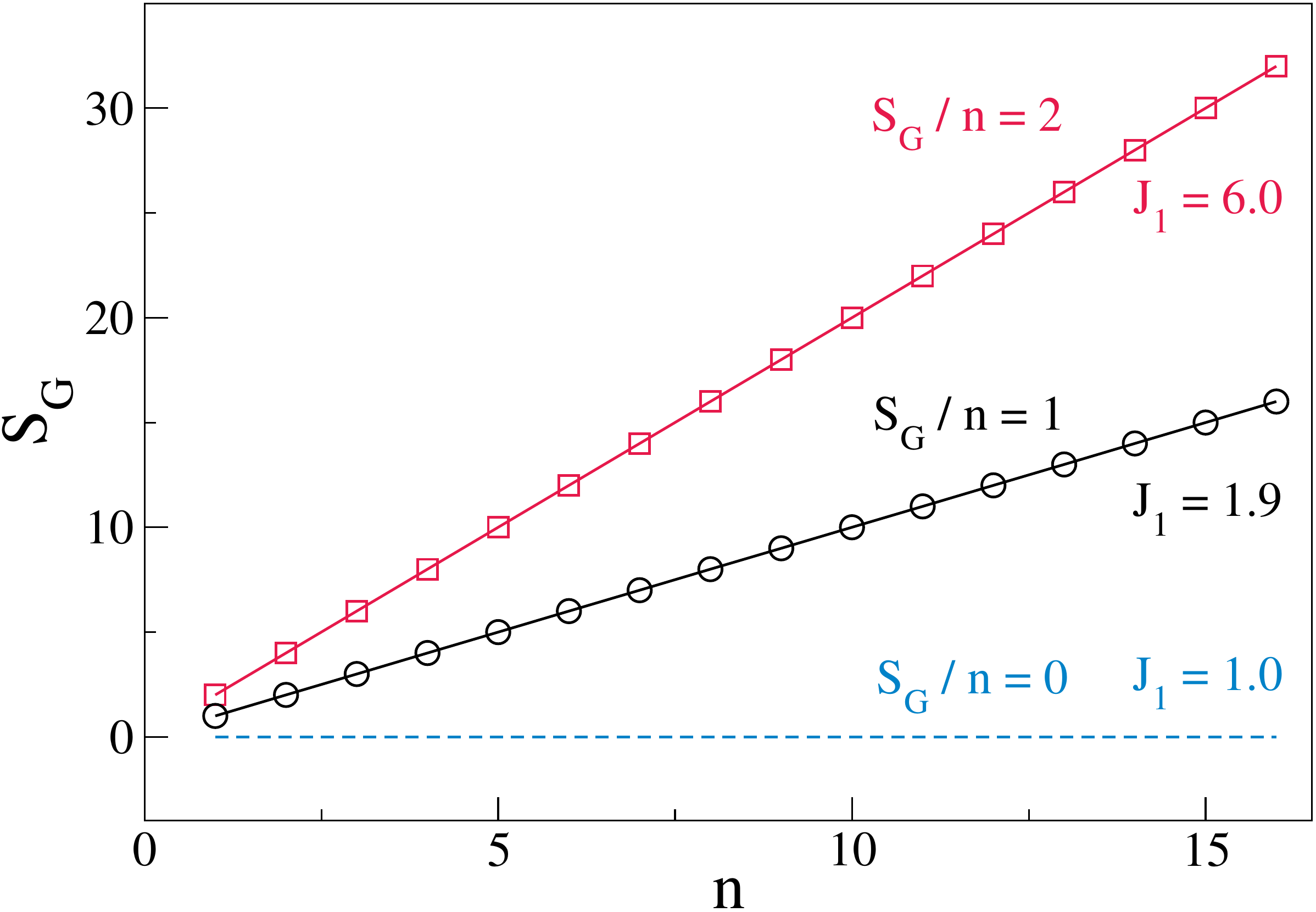}
	\caption{\label{fig:sg-n}The variation of the gs spin SG with 
	the number of unit cells $n$ is shown for $J_1 = 1.0$, 1.9 and 6.0.}
\end{figure}

We investigated the spin gaps at $J_1 = 5.1$ (where there is a 
peak in the $\Gamma_l$ values for the $N = 24$ spin system) as a function of 
system size from $N = 24$ to $N = 96$ spins, retaining 2400 block states in the 
finite DMRG calculations for both PBC and OBC. We find that for this $J_1$ value, 
the gaps $\Gamma_1$ to $\Gamma_n$ exhibit nonlinear variation with system size as 
shown in Figs.~\ref{fig:spgap}(c) and~\ref{fig:spgap}(d) for PBC and OBC, respectively. The 
convergence for higher values of $l$ in $\Gamma_l$ for PBC is poor and hence is not shown 
in Fig.~\ref{fig:spgap}(c). We find that for some system sizes (number of unit cells $n$) 
the excitation gaps $\Gamma_1$, $\Gamma_2$ and  $\Gamma_3$ vanish, for both PBC 
and OBC. We surmise that the vanishing of the gaps is because 
for these ``$n$'', the number of unit cells in the system is an integral multiple of the periodicity 
of  the spin-spin correlations. In region IV, the spin of the gs is $2n$, and indicates that all 
the ``free'' spins are ferromagnetically aligned. The transition from region III to 
region IV was followed for the 24 spin system under PBC, by varying $J_1$ in small 
increments. We find a step in the gs spin at an intermediate value $1 < S_G/n<2$ with 
a step width of 0.18 in $J_1$,  but this region could not be studied for larger system 
sizes due to convergence difficulties.  In region IV, with $J_1 > 5.63$, $S_G$ 
corresponds to twice the number of unit cells. We summarize the behavior of the gs 
in different regions, in Fig \ref{fig:sg-n} 

\subsection{Spin correlations}
To understand the spin structure in different regions of the parameter space, we
have studied the spin-spin correlations of the total spin 
$C(r)= \langle \vec{S}_k \cdot \vec{S}_{k+r} \rangle$, where
$k$ is the reference site of spins in the middle of the system. The $z$ component of the spin 
correlations, $C^{zz}(r) = \langle S^z_k S^z_{k+r}\rangle - \langle S^z_k 
\rangle \langle S^z_{k+r} \rangle$, shows behavior 
similar to the total spin correlation. The total spin correlations are shown for a 
system of 98 spins, which correspond to 12 unit cells for OBC. 
These are calculated in the gs with $S^z = S_G$. There are 
three different spin correlations that we have computed. They correspond to the
correlation between spins on the lower leg $C_1 (49, 49+2r)$, $C_2 (50, 50+2r)$ 
between spins on the upper leg and $C_3 (51, 51+4r)$ between ``free'' spins 
which reside on the lower leg. The reference site for the correlations is from 
the middle unit cell which for $C_1$ is site 49, for $C_2$ is site 50 and for $C_3$
is site 51. For convenience, we classify the spins on the lower leg as of two 
types, type 1 ``bound'' spins, which are bound to three nearest-neighbor spins 
and type 2 as ``free'' spins, which are the middle sites in the five and seven-membered rings. 

\begin{figure*}
\includegraphics[width=0.3\textwidth]{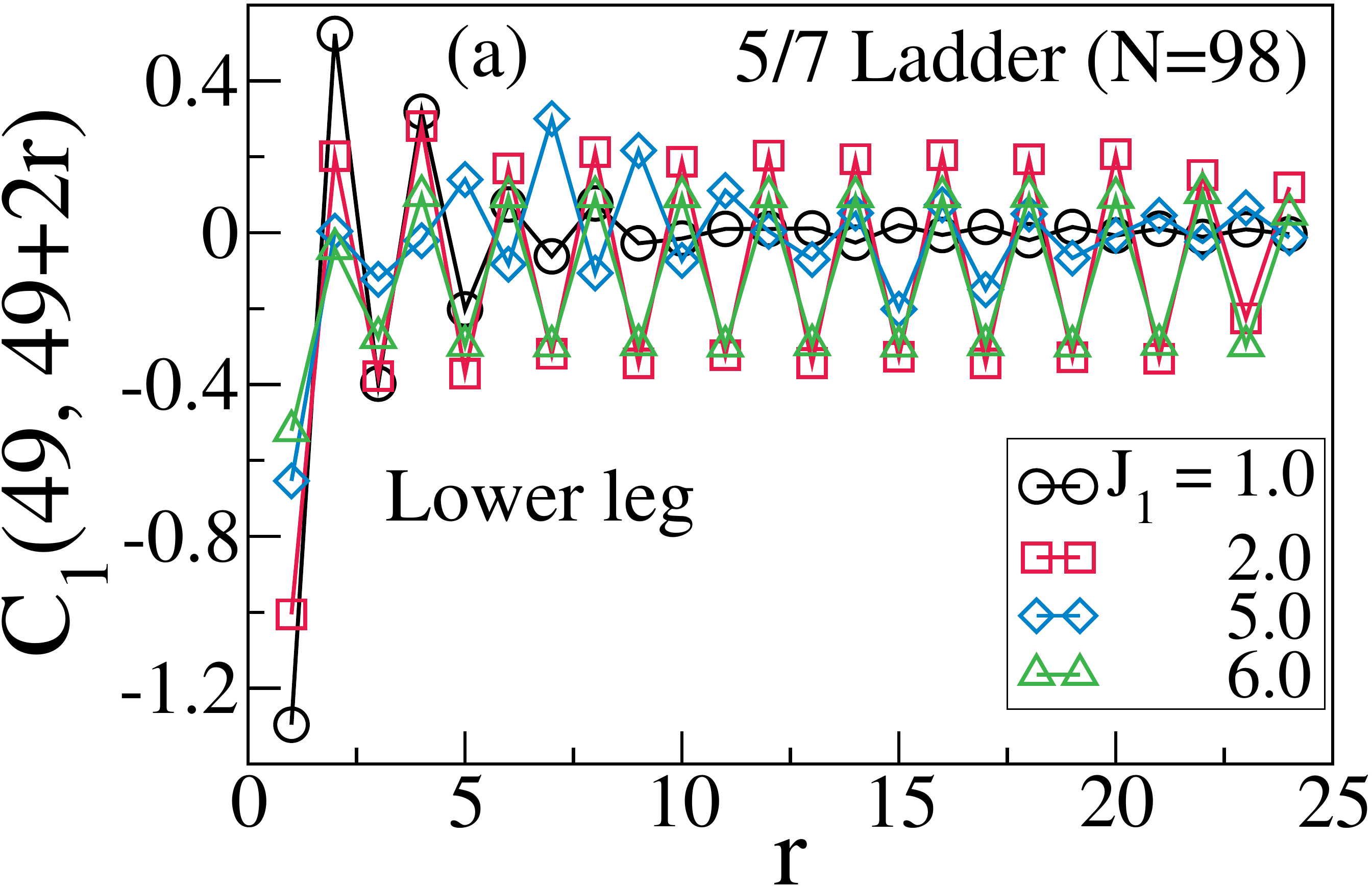}
\includegraphics[width=0.3\textwidth]{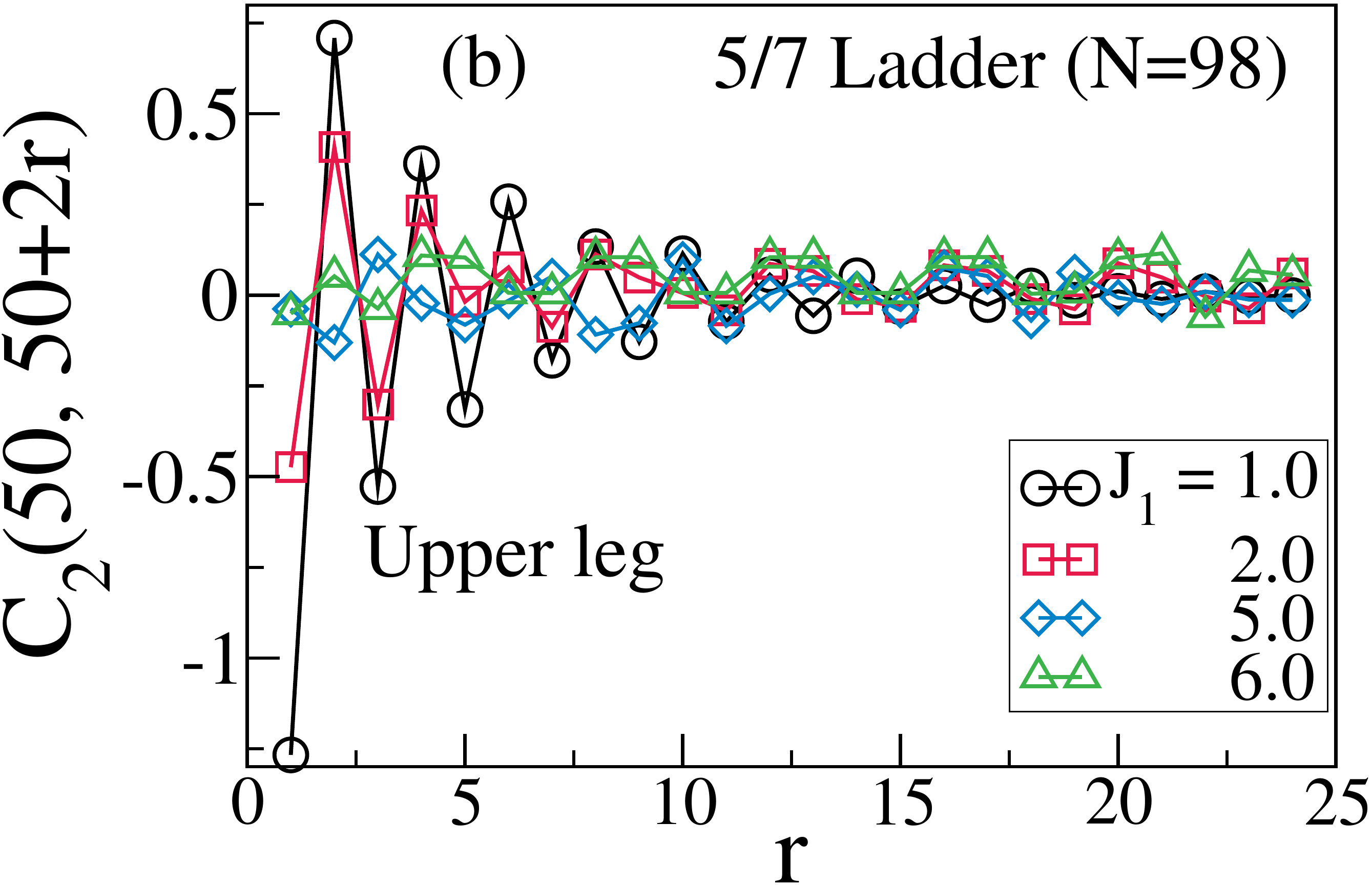}
\includegraphics[width=0.3\textwidth]{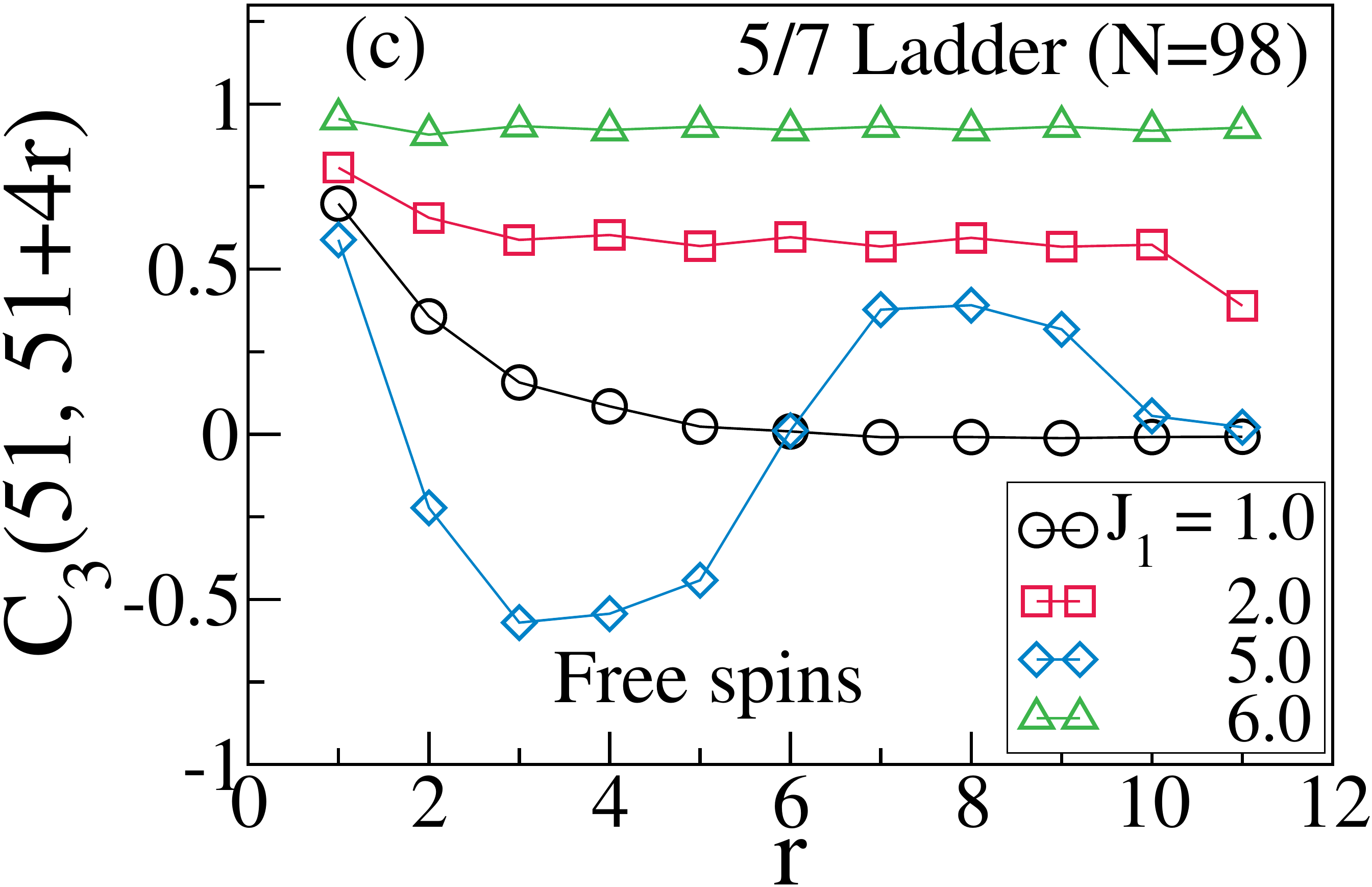}
\caption{\label{fig:corl}The spin-spin correlations between (a) spins in the
lower leg, (b) spins in the upper leg and (c) free spins at sites 3, 7, 11, etc.
for a 5/7 ladder with $N=98$ spins with OBC. Four $J_1$ values are chosen to
represent different regions of the phase diagram. $J_1 = 1.0$ for the nonmagnetic
phase, $J_1=2.0$ for the $S_G=n$ phase, $J_1=5.0$ for the reentrant phase, and
$J_1=6.0$ for the $S_G = 2n$ phase.}
\end{figure*}

In Fig.~\ref{fig:corl}, we show the spin correlations in the four different
regions of the parameter space. The correlations $C_1(r)$ shown in 
Fig.~\ref{fig:corl}(a) correspond to spins in the lower leg. 
The correlations are given from site 49 which is in the middle of the system.
For $J_1=0$, we have the correlation length of the Haldane system. We also 
find that as $J_1$ is increased, the correlation length gradually decreases
to $\sim 2.2$  for $J_1 = 1.0$. In the transition region between I and II, the 
correlations fluctuate rapidly and we can not extract a correlation length.
We note that for $J_1=1.0$, the system has a singlet gs. The 
correlations fall off rapidly and the correlation length 
$\xi$ is $\sim 2.2$ sites of the specified kind. In the spin-1 AF
chain the correlation length is longer by almost a factor of three and is 
approximately six sites.  The shorter correlation length can perhaps be 
attributed to the frustration in exchange interactions in the rings.
For $J_1 = 2.0$ and 6.0, the system is in a magnetic state, $C_1(r)$ decays 
very slowly and is AF in nature. In the reentrant phase the system 
goes to a nonmagnetic state and the spin correlations between the  spins
on the lower leg have long wavelength spin oscillations whose amplitude shows an 
exponential decay and corresponds to a noncollinear spin arrangement. From the 
spin correlations, it appears that the magnetic unit cell is tripled in this 
region. In Fig.~\ref{fig:corl}(b), $C_2(r)$ in the upper leg are shown for the four 
phases, with the reference spin being the 50$^{\rm{th}}$ spin in the system.  
$C_2(r)$ for both $J_1 = 1$ and 2 are AF and exponentially decaying 
with correlation length, $\xi \approx 3$. For $J_1 = 5$ and 6, $C_2(r)$ are 
vanishingly small, the magnitude is less than $\approx 0.05$ even for nearest-neighbor 
pair, and show long wavelength behavior. However, the amplitude of this 
wave is too small to definitively conclude this oscillatory behavior. The small correlation in upper leg 
at high $J_1$ is due to the strong dimer formation along the rungs and between 
($i$, 6) and ($i$, 8) sites. The correlations between ``free'' spins $C_3(r)$ shows a rapid 
decay in the nonmagnetic state at $J_1=1$, while those for $J_1 = 2$ and $J_1=6$, 
the correlations are ferromagnetic. For the $J_1=6$, the spins at these sites 
have almost completely aligned ferromagnetically, while for $J_1 =2$, the 
alignment is partially ferromagnetic.  This reflects in the net spin of the 
gs which is $2n$ for the $J_1=6$ case and $n$ for the $J_1=2$ case. In 
the reentrant phase, the free spins in each unit cell are aligned 
ferromagnetically while the alignment of these spins across unit 
cells is AF, with large periodicity. 

In summary, all the nearest-neighbor spin correlations are always AF, in the
nonmagnetic gs for $J_1$ in regime I, and the correlation lengths are
much shorter than the Haldane chain. For $J_1$ values in regime-II where 
$S_G \sim n$, the correlations in the lower leg  are
AF and very long ranged while those in the upper leg are
AF and fall off rapidly. The ``free'' spin correlations are
ferromagnetic with an amplitude of about $0.6$ and show very slow decay. When the
$J_1$ value is in the reentrant regime III, the lower leg  spin
correlations show formation of wave packets over approximately three unit cells.
The ``free'' spin correlations show a long wavelength oscillatory behavior with about
five unit cell wavelength, which corresponds to a long period N\'eel arrangement
of ``free'' spins. While in the classical, frustrated $J_1 - J_2$ model, the pitch
angle is dependent on $J_1 / J_2$, in the skewed ladder we have not been able to 
obtain a similar relationship. Besides, it is unlikely that a classical model
will exhibit a reentrant phase. In regime IV, the ``bound'' spins on the lower leg are
antiferromagnetically aligned and the correlations fall off very slowly with
distance. The correlations between spins on the upper leg show weak long
period N\'eel structure. The ``free'' spins are aligned ferromagnetically with very
long correlation length.

\subsection{Spin densities and bond order}
\begin{figure}
\includegraphics[width=3.4 in]{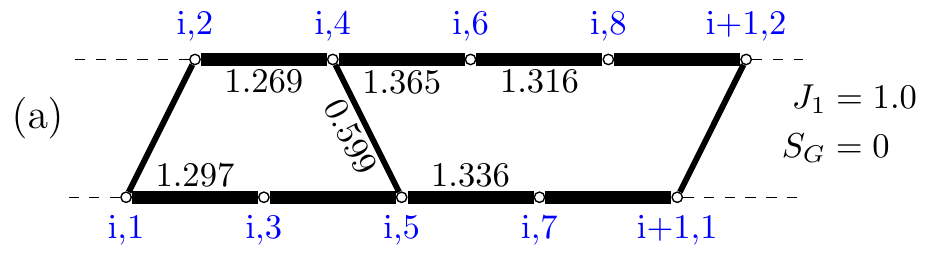}
\includegraphics[width=3.4 in]{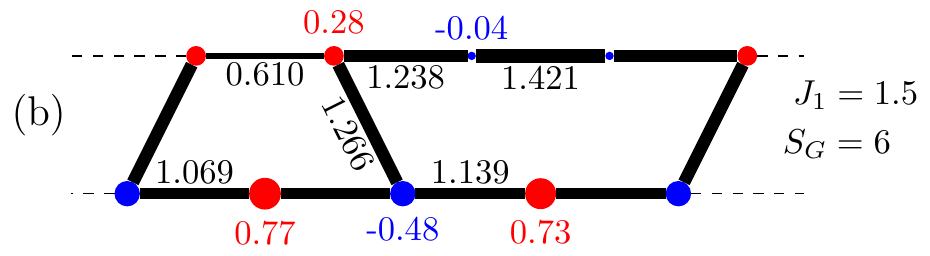}
\includegraphics[width=3.4 in]{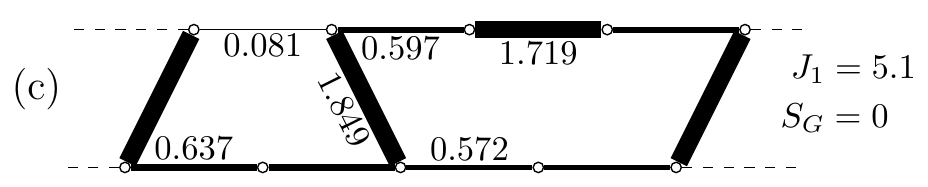}
\includegraphics[width=3.4 in]{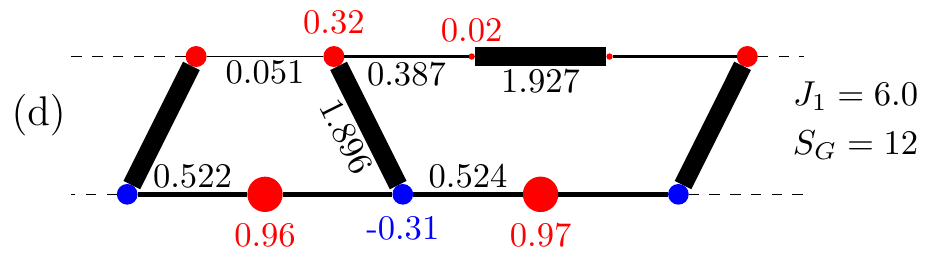}
	\caption{\label{fig:bo} Bond orders for 5/7 skewed ladder with $N=48$ spins
and PBC in the four different regions: (a)
	Nonmagnetic at $J_1=1.0$, (b) $S_G = n$ at $J_1 = 1.5$, (c) reentrant 
	nonmagnetic at $J_1 = 5.1$, and (d) $S_G = 2n$ at
$J_1 = 6.0$. The numbers adjacent to the bonds are the bond orders. 
	The width of the bonds is proportional to the magnitude
	of the bond order. The site numbering is given in 
	(a), the zero spin density is represented by open circles in (a) and (c),  the
	spin density in (b) and (d) is proportional to the area of the filled circles. 
	Red circles represent positive spin densities and blue circles represent
	negative spin densities.}
\end{figure}

The correlation lengths in the system are short, often less than distance to a
third equivalent nearest-neighbors. Hence, we can get qualitatively correct
behavior of the system in the thermodynamic limit from high accuracy DMRG
studies on a system with three unit cells. We have carried out studies on a 48
site spin-1 system, with PBC corresponding to six
unit cells $(n=6)$. We have retained 2400 block states for high accuracy in our DMRG computations.
We have computed the spin densities in the gs for 
$S^z = S_G$ and the bond orders of all the nearest-neighbor bonds. The system 
has reflection symmetry and hence there are five unique bonds 
and five unique sites. The spin densities are computed 
as the expectation value of $\langle \psi_{gs} \mid S_i^z \mid \psi_{gs} \rangle$. 
They are uniformly zero in the singlet gs. The bond orders are computed as  
$ b_{i,j} = -\langle \psi_{gs}\mid S_i \cdot S_j \mid \psi_{gs}\rangle$, where
$i$ and $j$ are nearest-neighbor bonds. When the gs of the system is a 
singlet (region I), from the bond orders (Fig.~\ref{fig:bo}) we can describe the 
system as weakly coupled spin-1 HAF chains. The upper leg has a  
BOW with a periodicity of four bonds, while the lower leg has a BOW with a
periodicity of two bonds. The rung bonds are weak and the leg bond orders vary between
$1.285$ and $1.370$. For comparison, in the spin-1 HAF, all the
bond orders are uniform and have a value of $1.40$. In region II, where
$S_G = n$, the upper leg bond between the sites in the pentagon become weak, the 
rung bonds become strong and the bonds in the lower leg also become slightly
weak. Besides, the bond on the upper leg between the sites which entirely belongs to the 
seven-membered ring also becomes strong. The rung bonds become much stronger 
while the ladder bonds become weaker.  
The spin densities at the ``free'' spin sites are nearly equal and there is a net 
negative spin density on the rung bonds with the spin density of the sites on 
the lower leg being large negative. The ``bonded'' spin sites on the seven-membered 
ring on the upper leg acquire small negative spin densities.  In 
region IV, where the $S_G = 2n$,  the rung bonds and the bond in the upper leg of 
the seven-membered ring almost form singlets, with a bond order close to $2.0$. All 
other bonds are very weak. The spin densities of the sites in the seven-membered 
ring which form the singlet are very nearly zero, while the rung bonds are 
qualitatively different with large negative (on the lower leg) and positive 
(on the upper leg) spin densities. The ``free'' spins are almost completely 
polarized and have spin densities that are very nearly unity. 

\subsection{Vector chirality}
Broken symmetry states give rise to different quantum phases whose properties
depend on the type of symmetry that is broken in the system. In general, broken
spatial inversion/reflection symmetry gives rise to bond the BOW phase,
whereas broken spin inversion symmetry gives rise to spin density wave (SDW). 
If both the spatial and spin inversion symmetries in the system are broken then 
the vector chiral phase arises and it leads to a spontaneous spin current in the 
system. For these symmetries to break simultaneously, the lowest energy states
in the two subspaces that the symmetry element divides the appropriate Hilbert 
space should be degenerate. In this case, any linear combination of the two 
low-lying states in the two subspaces which are even (odd) under both reflection 
and spin inversion will be degenerate resulting in symmetry breaking. The 
symmetry group of the 5/7 skewed ladder system consists of four elements: $E$, $P$, 
$\sigma$, and $\sigma P$, where $E$ is identity, $\sigma$ is the reflection symmetry,
and $P$ is the spin inversion symmetry and all these elements commute with each other leading 
to an Abelian group. The four irreducible representations correspond to $A^+$, $A^-$, 
$B^+$, and $B^-$. $A$ ($B$) corresponds to even (odd) under $\sigma$ while `$+$' (`$-$') 
corresponds to even (odd) under $P$. A BOW transition requires a degeneracy 
between the lowest states in $A^+$ and $B^+$ (or $A^-$ and $B^-$) subspaces.  Similarly 
an SDW transition requires a degeneracy of the lowest energy states in $A^+$ and $A^-$ 
(or $B^+$ and $B^-$) subspaces. For a vector chiral transition, the lowest energy 
states in $A^+$ and $B^-$ (or $A^-$ and $B^+$) subspaces must be degenerate.

\begin{table*}
\caption{\label{tab1}Two lowest energy levels from different $S^z$ sectors at specified
	$J_1$ values (see Fig.~\ref{fig:gsigma}).}
\begin{ruledtabular}
\begin{tabular}{ c  c  c  c  c  c }
	$J_1$ & $E(S^z = 0)$ & $E(S^z = 1)$ & $E(S^z = 2)$ & $E(S^z = 3)$ & $E(S^z = 4)$ \\ \hline
	1.07  & $-23.8471$   & $-23.8470$                                                \\
	      & $-23.8470$                                                               \\  \hline
        1.408 & $-25.0575$   & $-25.0575$   & $-25.0574$                                 \\
	      & $-25.0574$   & $-25.0574$                                                \\  \hline
	4.601 & $-44.7871$   & $-44.7871$   & $-44.7871$   & $-44.7871$                  \\
              & $-44.7871$   & $-44.7871$   & $-44.7871$                                 \\  \hline
	5.55  & $-51.7948$   & $-51.7948$   & $-51.7948$   & $-51.7948$   & $-51.7946$   \\
	      & $-51.7946$   & $-51.7946$   & $-51.7946$   & $-51.7946$                  \\
\end{tabular}
\end{ruledtabular}
\end{table*}
\begin{figure}
\includegraphics[width=3.2 in]{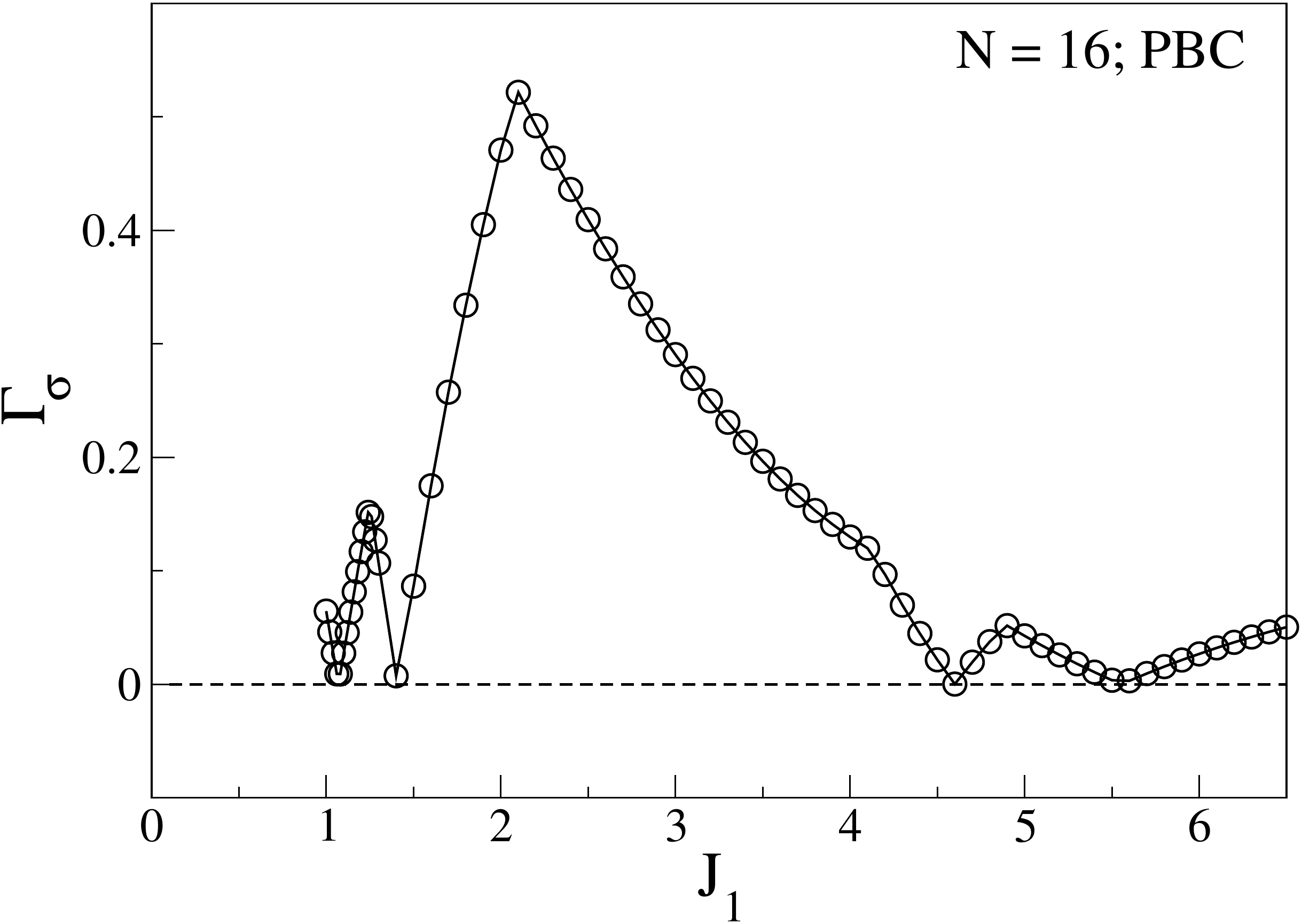}
	\caption{\label{fig:gsigma}The energy gap $\Gamma_{\sigma}$ (see Eq.~\ref{eq:gsig}) 
	between two lowest energy levels belonging to different reflection symmetry 
	subspaces. $\Gamma_{\sigma}$ vanishes at $J_1 = 1.07$, $1.408$, $4.601$ and $5.55$
	indicating degeneracies at these $J_1$ values.}
\end{figure}
Since the spin inversion symmetry, $P$ divides the $S^z = 0$ subspace into even and odd
total spin ($S$) sectors, the lowest energy states of the odd and even 
subspaces under $P$ should be degenerate to break the spin inversion symmetry. 
To determine the degeneracy of the lowest energy state in the odd and even 
subspaces under spin inversion symmetry, instead of employing `$P$' to divide the 
Hilbert space with $S^z = 0$ into even and odd total spin subspaces, we use the 
following argument. Whenever there is a degeneracy of the lowest energy states 
with odd and even total spin sectors, then the spin inversion symmetry is broken. 
We recognize the degeneracy of the gs when two states in the $S^z = 0$ sector 
are degenerate. In this case, we compute the energies of the lowest states 
in the higher $S^z$ sectors. The spin of the degenerate spin states is determined 
by following the degeneracies of the states in these sectors.
 
We calculate the energy gap $\Gamma_{\sigma}$, as the modulus of the difference 
in energy between the lowest energy states in
the $A$ and $B$ subspaces,
\begin{equation}
\Gamma_{\sigma} = |E_0 (\sigma=-1) - E_0 (\sigma=+1)|,
\label{eq:gsig}
\end{equation}
where $E_0 (\sigma=+1)$ and $E_0 (\sigma=-1)$ are the lowest energies in the 
even and odd subspaces under $\sigma$. In Fig.~\ref{fig:gsigma}, 
$\Gamma_{\sigma}$ is shown as a function of $J_1$ for a system size $N=16$ with 
PBC. We see that it vanishes at four values of $J_1$, 
namely $J_1=1.07$, $J_1=1.408$, $J_1=4.601$, and $J_1=5.550$.  At these values 
of $J_1$, we compute $\Gamma_l$ and find the two degenerate gs have spins $S=0$ 
and 1, $S=1$ and 2,  $S=2$ and 3, and $S=3$ and 4 respectively, at $J_1$ 1.07, 
1.408, 4.6, and 5.550. In Table~\ref{tab1} we show the lowest energy states in 
different $S^z$ sectors at the four points at which $\Gamma_{\sigma}$ vanishes. 
The degeneracy under reflection and spin inversion at a given $J_1$ implies a 
vector chiral state and nonzero spin currents for these $J_1$ values.

Usually, in all known systems, the chiral phase emerges either due to exchange 
anisotropy or due to an external magnetic field on a ferrimagnetic gs. 
However, in our system, due to the peculiar nature of the frustrated exchange 
interactions, accidental degeneracy occurs between the lowest energy states in 
the $A^+$ and $B^-$ symmetries. In the basis of these degenerate states, the 
spin chirality operator has nonzero eigenvalues. This leads to nonzero spin 
current in a well defined $S_z$ state, but the total spin is no longer conserved 
in the eigenstate. This implies spontaneous symmetry breaking, i.e., the 
eigenstate does not exhibit the full symmetry of the Hamiltonian.

The magnitude of the $z$ component of the spin current, $\kappa_z(j,k)$, is 
given by the eigenvalues of the matrix of the spin current operator for the $j-k$ bond, viz., 
$(\vec{S}_j \times \vec{S}_k)_z$ in the $\mid \psi_G(\pm) \rangle$ basis. The 
matrix, in this basis, is given by,
\[
\begin{pmatrix}
	\langle \psi(+) \mid \hat{\kappa}_z \mid \psi(+)\rangle &  \langle \psi(+) \mid \hat{\kappa}_z \mid \psi(-)\rangle \\
	\langle \psi(-) \mid \hat{\kappa}_z \mid \psi(+)\rangle  &  \langle \psi(-) \mid \hat{\kappa}_z \mid \psi(-) \rangle
\end{pmatrix},
\]
where the function $\mid \psi_G(+) \rangle$ is the lowest energy state in the 
even subspace for reflection and even subspace for spin inversion, similarly, 
$\mid \psi_G(-) \rangle$ is in the odd supspace under both symmetries.
The matrix elements of the spin current operator can be evaluated easily by using 
the operator identity 
\begin{eqnarray}
\kappa_z(j, k) &=& -i (\vec{S}_j \times \vec{S}_k)_z \
 \nonumber \\ 
&=& \frac{1}{2} (S^+_j S_k^- - S_j^-S_k^+ )
\label{eq:spcurrent}
\end{eqnarray}
The diagonal matrix element in the $2\times2$ matrix is zero, and the 
eigenvalues of the spin current are given by 
$\pm \frac{1}{2}\left| \langle \psi_G(+) \mid (S^+_j S_k^- - S_j^-S_k^+ ) \mid \psi_G(-) \rangle\right|.$
\vspace{2mm}

\begin{figure}
\includegraphics[width=3.4 in]{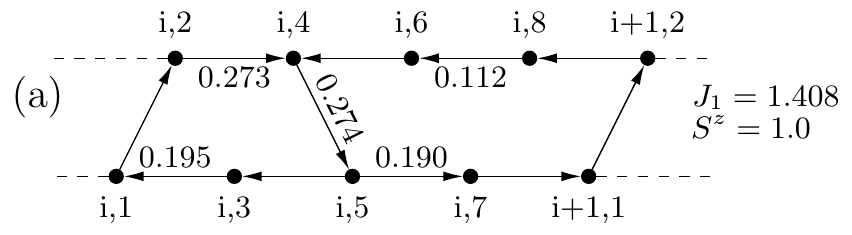}
\includegraphics[width=3.4 in]{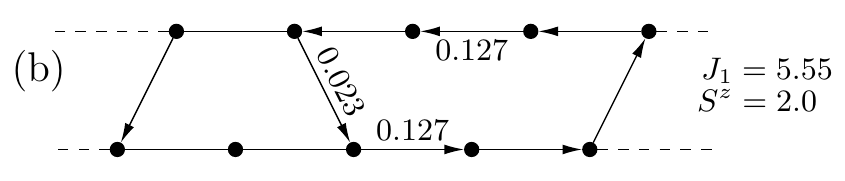}
\includegraphics[width=3.4 in]{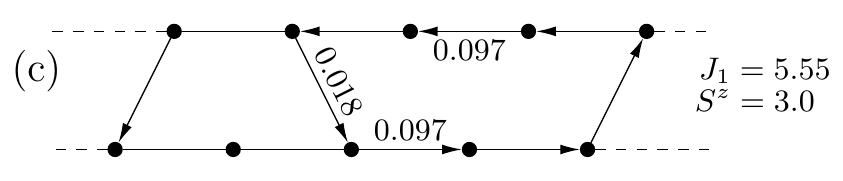}
	\caption{\label{fig:current}Spin currents $\kappa_z(j, k)$ of a 5/7 skewed 
	ladder of $N=16$ spins (see Eq.~\ref{eq:spcurrent}) for (a) $J_1 = 1.408$ in the $S^z = 1$ sector, 
	(b) $J_1 = 5.55$ in the $S^z = 2$ sector, and (c) $J_1 = 5.55$ in the $S^z = 3$ sector.
	Arrows indicate the direction of the current and the magnitude of the currents
	is given adjacent to the arrows. No arrowhead for a bond means the spin current is zero
	along that bond.}
\end{figure}

In Fig.~\ref{fig:current}, we show the spin currents for $J_1=1.408$ and $5.550$
for different degenerate $S^z$ values. Spin currents for all $J_1$
values, at which $\Gamma_{\sigma}=0$ and there is a degeneracy between states
of odd and even total spin, are shown in Table~\ref{tab2} for different $S^z$ values.
We find that the spin currents are large
for $J_1=1.408$ compared with $J_1=5.550$. At $J_1 = 1.408$, the spin current is
larger in the five-membered ring compared with that in the seven-membered ring.
Also, the direction of spin currents is opposite in the two rings. 
The spin currents are also not uniform for all the bonds, implying the mean angle between 
the spins depends upon the bond, as the spin current of a bond is a measure of 
the angle between orientations of the spins of the bond. The rung bonds have smaller
currents than the leg bonds and in the five-membered ring, the upper leg has
larger currents than the lower leg, while it is the opposite in the seven-membered 
ring. The spin current in the five-membered ring almost vanishes at
$J_1=5.550$ while it is much weaker in the seven-membered ring. The spin
currents of the rung bonds are very small and the spin current on the leg bonds
in the seven-membered ring becomes uniform. There is also weak dependence of the 
spin current on the $S^z$ value of the state for which it is calculated.
\begin{table*}
\caption{\label{tab2}Spin currents for different $J_1$ values in different $S^z$ 
	sectors. At $J_1 = 1.07$, $S=0$ and $S=1$ are degenerate. At $J_1 = 1.408$, 
	$S=1$ and $S=2$ are degenerate. At $J_1 = 4.601$, $S=2$ and $S=3$ are degenerate
	and finally at $J_1 = 5.5$, $S=4$ and $S=5$ are degenerate. $\kappa_z$ is 
	defined in Eq.~\ref{eq:spcurrent}}
\begin{ruledtabular}
\begin{tabular}{ c  c  r  r  r  r  r  r }
	$J_1$ & $S^z$ & $\kappa_z(1,2)$ & $\kappa_z(4,5)$ & $\kappa_z(1,3)$ & $\kappa_z(5,7)$ & $\kappa_z(2,4)$ & $\kappa_z(4,6)$ \\ \hline
	1.07  & 0 & $-0.302$ & $-0.302$ &  $0.165$ & $-0.158$ & $-0.237$ & $0.086$ \\ \hline
	1.408 & 0 & $-0.316$ & $-0.316$ &  $0.225$ & $-0.220$ & $-0.315$ & $0.129$ \\
	      & 1 & $ 0.274$ & $0.274$ &  $-0.195$ & $ 0.190$ & $ 0.273$ & $-0.112$ \\ \hline
	4.601 & 0 & $-0.045$ & $0.045$ &  $0$ & $0.205$ & $0$ & $-0.205$ \\
	      & 1 & $-0.042$ & $0.042$ &  $0$ & $0.193$ & $0$ & $-0.193$ \\
	      & 2 & $ 0.033$ & $-0.033$ &  $0$ & $-0.153$ & $0$ & $0.153$ \\ \hline
	5.55  & 0 & $ 0.027$ & $-0.027$ &  $0$ & $-0.147$ & $0$ & $0.147$ \\
	      & 1 & $-0.026$ & $ 0.026$ &  $0$ & $ 0.143$ & $0$ & $-0.143$ \\
	      & 2 & $-0.023$ & $ 0.023$ &  $0$ & $ 0.127$ & $0$ & $-0.127$ \\
	      & 3 & $-0.018$ & $ 0.018$ &  $0$ & $ 0.097$ & $0$ & $-0.097$ \\
\end{tabular}
\end{ruledtabular}
\end{table*}

\section{\label{sec4}Summary and Conclusion}
In this paper, we study quantum phases of a spin-1 HAF
model on a 5/7 ladder shown in Fig.~\ref{fig:schematic}(b). This 
system goes from a nonmagnetic state to a partially magnetized state for 
$J_1 > 1.06$, and for $J_1 > 1.44$ magnetization per unit cell is 
$m = S_G / n = 1$. The gs again goes to a nonmagnetic state for
$4.74 < J_1 <5.44$, and spins have noncollinear arrangement in this phase. For
large $J_1>5.63$, the gs goes to a magnetic state with $m = 2$.

The correlation length  in the 5/7 ladder decreases monotonically with $J_1$ and 
for $J_1=1$ the correlation length $\xi \sim 3$ lattice units in the singlet 
gs. The bond order of the rung bonds increase
monotonically with $J_1$, and in the large $J_1$ limit, the gs is a product of rung 
dimers and free spins at 3 and 7 sites in each unit cell. The uniform VBS state of spin-1 
chain disappears even for the small value of $J_1$ as spin-1/2 at the edges of the 
ladder gets pinned by the rung interaction and forms singlet pairs. In 
large $J_1$ limit, the spins at sites $8i-2$ and $8i$ ($i$ is the unit cell index)
form a strong singlet dimer, which is comparable to a spin-1 singlet dimer with 
bond order $\sim 2.0$. Thus, in the large $J_1$ limit we have a VB state with 
singlets on the rung bonds and between sites $6$ and $8$ in each unit cell. 
The free spins at sites $8i+3$ and $8i+7$ have ferromagnetic alignment.

For $J_1 < 1.06$, $\Gamma_{\sigma}$ vanishes and reflection symmetry is broken 
leading to dimer order in the system. For larger $J_1$ values, the gs is in 
ferrimagnetic state and for some $J_1$ values the lowest energy states in the 
even-even and odd-odd subspace under reflection and spin inversion become 
degenerate. This leads to both  inversion and spin-parity symmetry being
broken at the degeneracy points; therefore, the gs at this $J_1$ value 
possesses vector chirality and there is spontaneous  
spin current in the ladder system. This is unique as it can have both 
finite magnetization and spin current in the absence of an external magnetic
field. In this system, maximum gs magnetization per unit cell,
$m = 2$, while, for a spin-1/2 system
$m = 1$~\cite{geet, 57plateau}. In the spin-1/2 system magnetic moments are 
localized mostly on ($4i-1$) sites of the system and other sites have 
vanishingly small spin density~\cite{geet, 57plateau}. However, in the spin-1 system while
magnetization contribution comes mostly from sites ($4i-1$), there is
antiferromagnetically aligned spin density in the lower leg and
ferromagnetically aligned spins in the upper leg due to strong singlet dimer
formation along the rung. The isolated singlet spin-1 dimer between sites $8i-2$
and $8i$ in this extended system is unique.  

In conclusion, the HAF spin-1 system on a skewed 5/7 ladder is unique with different
ferrimagnetic gs, and this system exhibits a plethora of exotic phases in the gs on
tuning $J_1$.  The spin arrangements of the spin-1 system are vastly different from
those on the spin-1/2 on this lattice. This is a unique ladder system where
a singlet spin-1 dimer and BOW can coexist. The topological
phase of the spin-1 chain vanishes for any finite value of $J_1$. The HAF 5/7 ladder
system can be mapped onto a spin-1 chain with an AF nearest- and next-nearest-neighbor 
interaction $J_1$ and $J_2$ with periodically missing $J_1$ bonds. This 
system may be realized in molecular magnets based on transition metal compounds.

\begin{acknowledgments}
MK thanks Department  of Science and Technology (DST), India
for Ramanujan fellowship. SR thanks Indian National Science Academy and
DST-SERB for supporting this work.

SD and DD contributed equally to this work.
\end{acknowledgments}

\bibliography{ref57sp1}
\end{document}